\documentclass[twocolumn,preprintnumbers,amsmath,amssymb,prl,superscriptaddress]{revtex4}

\usepackage{graphicx}
\usepackage{dcolumn}
\usepackage{bm}
\usepackage{subfigure}
\begin{document}

\title{Excluded volume effects on semiflexible ring polymers}

\author{Fabian  Drube}
\affiliation{Arnold Sommerfeld Center for Theoretical Physics and Center for NanoScience, Department of Physics,\\ Ludwig-Maximilians-Universit\"at M\"unchen, Theresienstra\ss e 37, D-80333 M\"unchen, Germany
}
\author{Karen Alim}
\affiliation{Arnold Sommerfeld Center for Theoretical Physics and Center for NanoScience, Department of Physics,\\ Ludwig-Maximilians-Universit\"at M\"unchen, Theresienstra\ss e 37, D-80333 M\"unchen, Germany
}
\author{Guillaume Witz}
\affiliation{Laboratoire de Physique de la Mati\`ere Vivante, Ecole Polytechnique F\'ed\'erale de Lausanne (EPFL), CH-1015 Lausanne, Switzerland}
\author{Giovanni Dietler}
\affiliation{Laboratoire de Physique de la Mati\`ere Vivante, Ecole Polytechnique F\'ed\'erale de Lausanne (EPFL), CH-1015 Lausanne, Switzerland}
\author{Erwin Frey}
\affiliation{Arnold Sommerfeld Center for Theoretical Physics and Center for NanoScience, Department of Physics,\\ Ludwig-Maximilians-Universit\"at M\"unchen, Theresienstra\ss e 37, D-80333 M\"unchen, Germany
}

\date{\today}

\begin{abstract}
Two-dimensional semiflexible polymer rings are studied both by imaging circular DNA adsorbed on a mica surface and by Monte Carlo simulations of  phantom polymers as well as of polymers with finite thickness. Comparison of size and shape of the different models over the full range of flexibilities shows that excluded volume caused by finite thickness induces an anisotropic increase of the main axes of the conformations, a change of shape, accomplished by an enhanced correlation along the contour. 
\end{abstract}

\keywords{Monte Carlo simulation, semiflexible polymer, DNA, statistical mechanics, conformation,polymers on surfaces, biofilaments}
\maketitle

DNA is one of the most versatile building blocks for the self-assembly of nanoscale structures \cite{Seeman:2003p5957}. The complementary base-pairs enables almost unlimited possibilities in designing highly tailored constructs \cite{CHEN:1991p1437}. But even without specially designing  sequences for specific base-pairing, biopolymers such as DNA or cytoskeletal filaments self-assemble in their natural environment. DNA occurs both in linear and circular forms and condenses into toroidal structures \cite{Bloomfield:1991p5961,Hud:2001p5990}. Actin assembles into filaments and bundles \cite{FLodish:2008p6016} and also builds ring \cite{Tang:2001p1647,Limozin:2002p1622,Claessens:2006p142}, and racket like \cite{Cohen:2003p1614,Lau:2009p5614} complexes. From a geometrical perspective filaments denote the simplest building blocks. The next level of complexity is reached when topological constrained forms as in rings arise. Considering polymer rings as building blocks their size and shape are of eminent importance. Both size and shape strongly depend on two internal biopolymer properties: the ability to bend and the effective diameter of the polymer. Indeed the shape of polymer rings has been investigated theoretically regarding the influence of the polymers flexibility $L/l_p$, given by the ratio of total polymer length $L$ and its persistence length $l_p$ \cite{alim2007prl}. However, our experimental observations show that the previously neglected finite thickness not only regulates the absolute size  \cite{flory1953} of a polymer configuration but also its shape. This observation is especially important in confined geometries utilized in the preparation of biopolymer assemblies of higher order \cite{Lau:2009p5614}. 

\begin{figure}[b]
\centering
\includegraphics[width=0.45\textwidth]{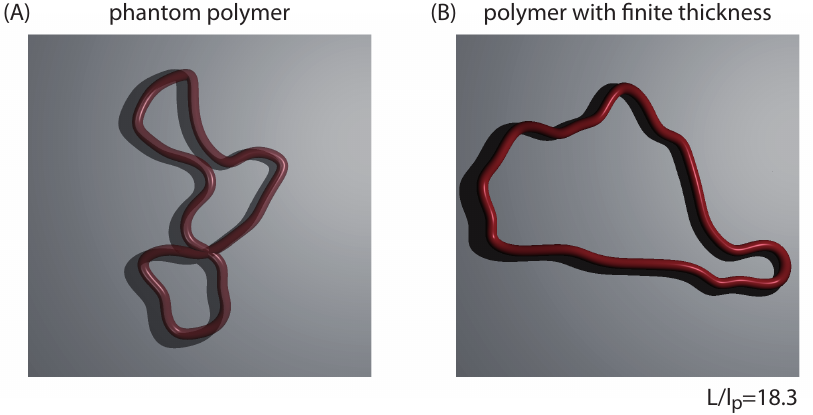}
\caption{Simulation snap-shot of a phantom polymer (A) and a polymer with finite thickness (B) on a plane. Phantom polymers bear overlaps and kink-like bending resulting in smaller configurations with more aspherical shapes than polymers with finite thickness.}
\label{fig_cartoon}
\end{figure}

Coarse grained polymer models rely on phantom chains, which allow segments to overlap. To describe real polymers with finite thickness, the excluded volume of a polymer chain is accurately accounted for by tube models \cite{rubinstein2003}, where the tube imposes a hard core potential, see Fig.~\ref{fig_cartoon}. To access the effects of finite thickness and topology experimentally in a well-defined setup, we investigate circular DNA adsorbed on a mica surface, which has previously been shown to obey two dimensional worm-like chain statistics \cite{rivetti1996, witz2008}. Our data verifies Flory's predicted growth in size due to polymer's finite thickness  \cite{flory1953}, as has also been accomplished experimentally in different contexts \cite{maier1999, valle2005, ercolini2007} and theoretically by self-consistent and renormalization group theories \cite{descloizeaux1990}. Since DNA can be produced in different lengths, it serves as a model system to investigate the shape over the full range of flexibility, which so far has only been forecasted theoretically for phantom polymers in three dimensions \cite{alim2007prl}.

Here we study the effects of excluded volume caused by finite thickness on semiflexible polymer rings by imaging circular DNA on a mica surface and performing Monte Carlo simulations of semiflexible phantom polymers and polymers with finite thickness. Calibrating the polymer's diameter $d$ by comparing the tangent-tangent correlation, we obtain good  quantitative agreement between experiments and  simulations with diameter estimates from polyelectrolyte theory. The finite thickness leads to an apparent stiffening and an increase of the principal axes of the polymers configurations as observed by the mean square radius of gyration.  In contrast to Flory's prediction, the asphericity as a shape measure proves the growth to be anisotropic, resulting in a change of shape.   

The DNA rings without superhelicity were produced from nicked plasmids pSH1, pBR322, and pUC19 with flexibilities of $L/l_p=40$, $30$, and $18.3$, respectively. Plasmid pUC19 was treated with restriction enzyme RsaI to produce three different linear fragments, and using T4 DNA ligase, minicircles of different flexibilities were obtained $L/l_p=12.4$, $4.6$. In order to access the trajectory of the DNA rings by Atomic Force Microscopy, the sample was deposited on mica, see supplementary material for details. Previous analysis of us \cite{witz2008} proofed the tangent-tangent correlation and the mean square radius of gyration to obey two dimensional statistics of the polymer rings.

For the Monte Carlo simulation of a semiflexible polymer rings with persistence length $l_p$,  preceding simulation methods were customized to describe two dimensional polymers and extended to incorporate finite thickness as outlined in the supplementary material. In the simulations with finite thickness configurations including intersections of tubes of the diameter $d$ around each segment are rejected. Uncorrelated data sets are obtained by taking configurations every $10^6$ Monte Carlo steps for phantom polymers and every $10^8$ steps for simulations {with finite thickness}. Large ensembles are sampled such that the statistical error based on a normal distribution of the observable is of the size of the symbols in all figures shown.

Semiflexible polymers are well described by the wormlike chain model, where the polymer is  modeled as an elastic rod with bending modulus $\kappa$ \cite{rubinstein2003}. Representing the polymer by a  differential space curve $\mathbf{ r(s)}$ of length $L$ parametrized by an arc length $s$, its statistical properties are determined by the elastic energy $ \mathcal{H}=\kappa/2\,\int_0^L\,ds \left[\partial \mathbf{ t}(s)/\partial s \right]^2$, where $\mathbf{ t}(s) = \partial \mathbf{r}(s)/\partial s$ is the tangent vector. The persistence length $l_p$ as a measure of the stiffness is defined by the initial decay of the mean tangent-tangent correlation $\langle \mathbf{ t}(s)\mathbf{ t}(s')\rangle = \exp(-|s-s'|/l_p)$, given by $l_p=\frac{2\,\kappa}{k_B\,T}$  for a polymer embedded in two dimensions.

 Size and shape of a polymer are comprised in the radius of gyration tensor,
\begin{equation}
Q_{ij}=\frac{1}{L}\int ds\,\mathbf{r}_i(s)\mathbf{r}_j(s)-\frac{1}{L^2}\int ds\,\mathbf{r}_i(s)\int d\tilde{s}\,\mathbf{r}_j(\tilde{s})\,,
\end{equation}
whose eigenvalues $\lambda_1$  and $\lambda_2$ define the spatial extent of the polymer along its principal axes. The squared radius of gyration measures the total size of an object and hence is given by the sum of the two eigenvalues,
\begin{equation}
 R^2_g=\lambda_1+\lambda_2\,.
\end{equation}
 The criterion for the shape of a polymer is the asphericity, which is given by the normalized variance of $\lambda_1$ and $\lambda_2$  \cite{aronovitz1986}, yielding in two dimensions,
\begin{equation}
 \Delta=\frac{(\lambda_1-\lambda_2)^2}{(\lambda_1+\lambda_2)^2}\,.
 \label{asphericity}
\end{equation} 
The asphericity ranges between $0$ and $1$; $\Delta=0$  for the most spherical object in two dimensions, the ring, and $\Delta=1$ for the most aspherical configuration, a rod. 

\begin{figure}[t]
\centering
\includegraphics[width=0.35\textwidth]{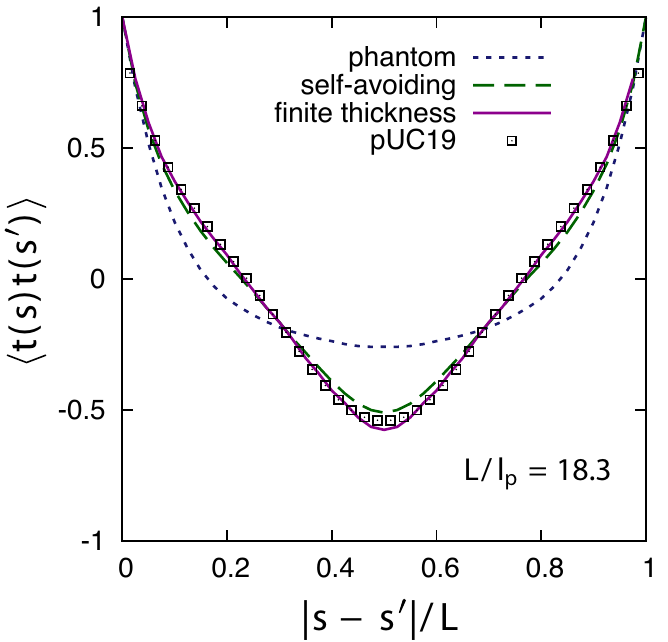}
\caption{Tangent-tangent correlation of a semiflexible polymer ring for different theoretical models and for DNA (pUC19). The finite thickness is calibrated to $d/l_p=0.13$. Incorporating the same persistence length models with higher self-exclusion show enhanced correlations.}
\label{fig_tt}
\end{figure}

To model real polymers, two internal parameters are required, the persistence length $l_p$  and the diameter $d$ of  the filament. For the persistence length of DNA we use the widely accepted value of $l_p = 50 nm$. The effective diameter of DNA, being a polyelectrolyte, changes in a predictable manner in response to its surrounding ionic solution, as it has been determined theoretically \cite{stigter1977} and experimentally \cite{rybenkov1993}. For our experimental conditions the ratio of diameter to persistence length is  $d/l_p=0.13$. To calibrate the simulation parameters to the experimental data, the tangent-tangent correlation is an ideal observable reflecting the statistics along the whole contour of the polymer. The agreement between simulation with finite thickness and experimental data is exemplified for the tangent-tangent correlation for a flexibility of $L/l_p=18.3$ considering plasmid pUC19 in Fig.~\ref{fig_tt}.
Included in the graph are also results for a phantom polymer, where self-intersecting configurations are permitted, and for reasons of comparison the trivial limit of vanishing thickness $d \rightarrow 0$ denoted self-avoiding polymer. In the latter case, only intersections of the polymer backbone are rejected. This limit does not describe the experimental data quantitatively as good as the simulations with a thickness of  $d/l_p=0.13$ ($7\%$ increase of the reduced chi-square). Hence, the diameter estimate based on polyelectrolyte theory yields good quantitative agreement.

In the tangent-tangent correlation function of  Fig.~\ref{fig_tt}, two effects are evidenced if comparison is made between a polymer with finite thickness and a phantom chain of same persistence length (solid and dashed line in  Fig.~\ref{fig_tt}, respectively), one due to the finite thickness of the polymer and the other to the circular topology. On short distances, the finite thickness restricts the available conformational space, thus increasing the directional correlation. In order to respect the circular topology, the correlation function must become more negative on distances $s\approx L/2$. In summary, the polymer with finite thickness appears effectively stiffer.
\begin{figure}[t]
\centering
\includegraphics[width=0.45\textwidth]{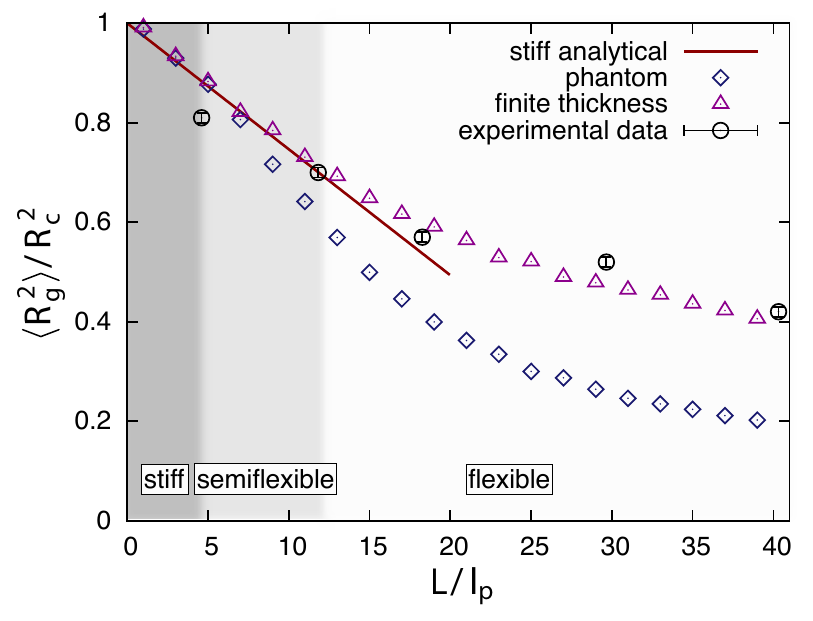}
\caption{Squared radius of gyration $\left<R^2_g\right>$ compared to the size of the corresponding rigid ring $R_c^2=(L/2\pi)^2$ vs. $L/l_p$. The influence of finite thickness arises in the semiflexible regime after an effective stiffening in accordance with analytical predictions for stiff rings. In the flexible regime, the increase in size is required to model the experimental data. Error bars indicate the statistical errors.}
\label{fig_rg}
\end{figure}

Similarly, this effective stiffening also affects the overall size of the polymer as measured by the squared radius of gyration $\left<R_g^2\right>$. In Fig.~\ref{fig_rg} $\left<R_g^2\right>$ normalized with the radius of gyration of a rigid circle, namely $\left<R_g^2\right>/R_c^2$, is plotted versus the flexibility parameter $L/l_p$ for the considered models together with the data. Three regimes are discerned; in the stiff regime all simulation data for the radius of gyration normalized by total polymer length versus flexibility $L/l_p$ follow the predicted linear decay \cite{alim2007epje}. The stiff regime extends up to high flexibilities compared to open chains. For phantom polymers this effect can be accounted for by an effective stiffening due to the topological constraint of a ring \cite{alim2007epje}. For polymers with finite thickness the squared radius of gyration follows the analytic result for the stiff limit up to even higher flexibilities of approximately $L/l_p=12$ indicating a further effective stiffening due to the polymer's thickness  i.e.~as a result of excluded volume. This is in agreement with the observation of enhanced correlations in the tangent-tangent correlation.
The semiflexible regime is a crossover region, where departing from the analytically determined stiff limit, phantom polymers show a linear decay as flexibility increases suggesting an initial step by step excitation of higher modes. In contrast, for polymers with finite thickness these initial higher modes are suppressed, resulting in a direct transition from the linear decay in the stiff limit to the power law decay in the flexible regime. Such polymers with finite thickness deviate from the stiff limit to larger sizes as is also observed for three-dimensional phantom polymer rings \cite{alim2007epje}. 
Finally, in the flexible regime both models have substantially different radii of gyration. In contrast to phantom polymers, polymers with finite thickness show notably larger sizes recovering Flory's swelling effect. The scaling exponent agrees with Flory's predicted exponent (data not shown). The segments of phantom polymers  overlap strongly to maximize entropy as flexibility permits. Precisely those modes are, however, forbidden for polymers with {finite thickness} yielding a larger mean squared radius of gyration. Flory's argument oversimplifies a semiflexible chain of segments to an ideal gas and assumes that all chain segments overlap with an equal probability with each other. This results in a growth in size that is equally large along all principal axes of the polymer. Hence, Flory's description predicts an {\it isotropic increase of the principal axes}, which will be tested below when considering the asphericity. The experimental data in Fig.~\ref{fig_rg} are in agreement with the finite thickness polymers, providing solid evidence that a description of DNA as a semiflexible polymer with persistence length $l_p  =50\ nm$ and effective diameter $d/l_p=1.3$ is a faithful description of DNA conformations. Indicated in the graph (Fig.~\ref{fig_rg}) are only statistical errors of a Gaussian distributed observable as a lower estimate of the statistical error. Furthermore, systematic errors may arise, first, due to the limited resolution of the AFM images, and second, due to the fact that the minicircles are not nicked and may thus experience a slight distortion.  

\begin{figure}[t]
\centering
\includegraphics[width=0.45\textwidth]{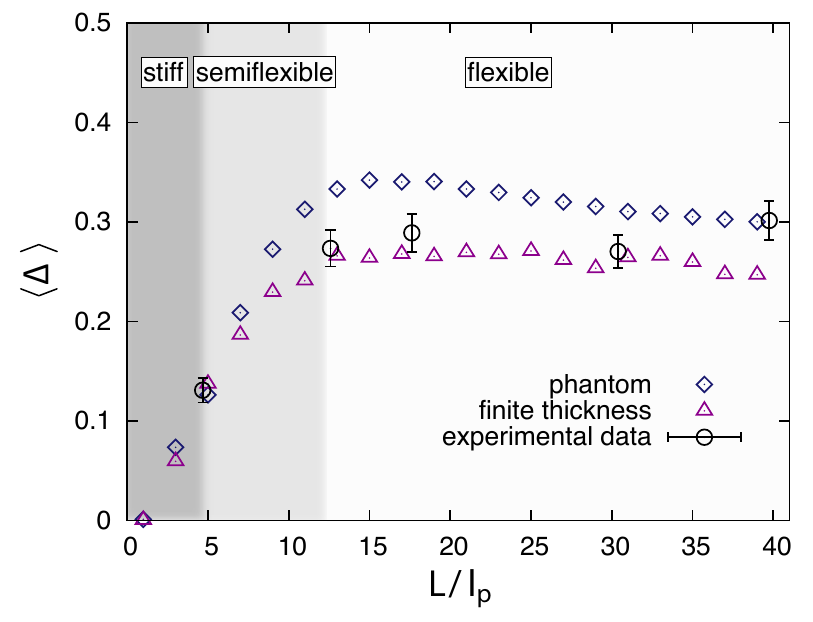}
\caption{The asphericities of the two models deviate in the semiflexible regime due to the influence of finite thickness. Through enhanced correlations between adjacent segments and prominent growth of the minor principal axis forced by pushing overlapping segments apart finite thickness makes the elliptical shape of ring polymers rounder. }\label{fig_asp}
\end{figure}

In order to test Flory's prediction, the asphericity $\left<\Delta\right>$ (Eq. \ref{asphericity}) is plotted in Fig.~\ref{fig_asp} as a function of the  flexibility parameter $L/l_p$ and the three regimes appear again. Starting from a ring configuration with $\Delta=0$ for infinite stiffness $L/l_p=0$, the asphericity grows linearly for both models in the stiff region due to the fact that polymers have an elliptical shape of increasing eccentricity, as it has been recently predicted by scaling arguments \cite{alim2007prl}. Self-exclusion plays no role in these configurations because the segments are well separated from each other due to the high bending energy of stiff polymers. The increase in asphericity is continued in the semiflexible regime until it starts to decrease slowly for flexible polymers, clarifying ambiguous simulations \cite{camacho1991,norman1992}.  For phantom polymers the asphericity decreases down to $\langle\Delta\rangle=0.2625$ in the Gaussian limit \cite{diehl1989}. The three regimes of the phantom polymer resemble results for three-dimensional phantom rings, which, however, show a more pronounced decrease in asphericity in the flexible regime as the third spatial dimension enables more compact configurations. 

Focussing on the magnitude of the asphericity in the two dimensional models of phantom polymers and polymers with finite thickness, we find that  the curves commence to deviate from each other in the semiflexible regime, showing fundamentally different values of asphericity in the flexible region. Hence, the ratio of the principal axes is strikingly different and the growth in size induced by the polymer's thickness is anisotropic along the principal axes yielding more spherical configurations compared to a phantom polymer. This is opposite to simulation results for open random walks in three dimensions, where self-avoidance has been found to lead to slightly more aspherical configurations \cite{cannon1991, bishop1988}. In fact in three dimensions, a random walk is rarely intersecting its own trajectory. Then our present work suggests the remaining governing impact of self-avoidance in three dimensions to be an effective stiffening. Such a stiffening yields more aspherical shapes for open semiflexible polymers in three dimensions.   

Confinement, like in the present case of two dimensional ring polymers, on the other hand, forces polymer segments to overlap much more frequently. Concerning two dimensional polymer rings, the notion of an aspherical shape indicates that one principal axis is much  
longer than the other like in an ellipse. In the apices of the ellipse, the segments are prone to overlap with neighboring segments on a local level, while segments in the convex part of the ellipse tend to overlap with segment being separated approximately half the total length along the contour. The finite thickness now effectively stiffens the polymer inducing less bending at the apices and it increases the minor principal axis by pushing segments in the convex region apart, see Fig.~\ref{fig_cartoon}. This results in a more spherical configuration for polymer rings with finite thickness, as observed in Fig.~\ref{fig_asp}. As the asphericity distribution is highly skewed our statistical errors underestimate the actual error justifying slight deviations between our sets of data. Considering two dimensional polymer rings the good agreement between simulations and experimental data over the broad range of flexibilities manifests the role of finite thickness and its effects of effective stiffening and anisotropic change in shape. 

In conclusion, we have analyzed the impact of excluded volume caused by the finite thickness of polymer rings in two dimensions over the whole range of flexibility, both by observing DNA rings on mica surface and by computer simulations of phantom and finite thickness polymers. We find that the experimental data can only be explained by the latter, where each segment of the polymer is represented by an impenetrable tube. From the comparison of the different models, we determine two effects of finite thickness. Firstly, tangent-tangent correlation shows an enhanced correlation due to the steric constraints of the neighboring segments, leading to an effective stiffening observed in the semiflexible regime of the mean-square radius of gyration. Secondly, in the flexible regime Flory's swelling is recovered. However, the asphericity discloses an anisotropic change in shape.  Manifesting these properties should enable a new understanding of the conformation statistics of biopolymers such as DNA and F-actin. A basis on which biopolymer assemblies can be designed to develop new nanomaterials.

F.D., K.A.~and E.F.~acknowledge financial support of the German Excellence Initiative via the program "Nanosystems Initiative Munich (NIM)", of the DFG through SFB 486, and of the LMUinnovativ project "Functional Nanosystems (FuNS)". K.A.~acknowledges funding by the Studienstiftung des deutschen Volkes. G.W.~and G.D.~acknowledge the support by the Swiss National Science Foundation through grants Nr.~200020-116515 and 200020-125159.

\providecommand*{\mcitethebibliography}{\thebibliography}
\csname @ifundefined\endcsname{endmcitethebibliography}
{\let\endmcitethebibliography\endthebibliography}{}

\end{document}